\documentclass[twocolumn]{aastex62}

\graphicspath{{./}{figures/}}
\submitjournal{ApJ}
\shorttitle{GS 2000+25}
\shortauthors{Rodriguez et al.}
\begin{document}

\title{GS\,2000+25: The Least Luminous Black Hole X-ray Binary}

\correspondingauthor{Jennifer Rodriguez}
\email{rodri756@msu.edu}
\author[0000-0003-1560-001X]{Jennifer Rodriguez}
\affiliation{Center for Data Intensive and Time Domain Astronomy, Department of Physics and Astronomy, Michigan State University, East Lansing MI 48824, USA}

\author[0000-0003-1814-8620]{Ryan Urquhart}
\affiliation{Center for Data Intensive and Time Domain Astronomy, Department of Physics and Astronomy, Michigan State University, East Lansing MI 48824, USA}
\author[0000-0002-7092-0326]{Richard M. Plotkin}
\affiliation{Department of Physics, University of Nevada, 1664 N. Virginia St, Reno, Nevada, 89557, USA}
\author[0000-0001-8424-2848]{Teresa Panurach}
\affiliation{Center for Data Intensive and Time Domain Astronomy, Department of Physics and Astronomy, Michigan State University, East Lansing MI 48824, USA}
\author[0000-0002-8400-3705]{Laura Chomiuk}
\affiliation{Center for Data Intensive and Time Domain Astronomy, Department of Physics and Astronomy, Michigan State University, East Lansing MI 48824, USA}
\author[0000-0002-1468-9668]{Jay Strader}
\affiliation{Center for Data Intensive and Time Domain Astronomy, Department of Physics and Astronomy, Michigan State University, East Lansing MI 48824, USA}
\author[0000-0003-3124-2814]{James C.A. Miller-Jones}
\affiliation{International Centre for Radio Astronomy Research, Curtin University, GPO Box U1987, Perth, WA 6845, Australia}
\author[0000-0001-5802-6041]{Elena Gallo}
\affiliation{Department of Astronomy, University of Michigan, 1085 S University, Ann Arbor, MI 48109, USA}
\author[0000-0001-6682-916X]{Gregory R. Sivakoff}
\affiliation{Department of Physics, CCIS 4-183, University of Alberta, Edmonton, AB T6G 2E1, Canada}




\begin{abstract}

Little is known about the properties of the accretion flows and jets of the lowest-luminosity quiescent black holes. We report new, strictly simultaneous radio and X-ray observations of the nearby stellar-mass black hole X-ray binary GS~2000+25 in its quiescent state. In deep \emph{Chandra} observations we detect the system at a faint X-ray luminosity of $L_X = 1.1^{+1.0}_{-0.7} \times 10^{30}\,(d/2 {\rm \,\, kpc})^2$ erg s$^{-1}$ (1--10 keV). This is the lowest X-ray luminosity yet observed for a quiescent black hole X-ray binary, corresponding to an Eddington ratio $L_X/L_{\rm Edd} \sim 10^{-9}$. In 15 hours of observations with the Karl G. Jansky Very Large Array, no radio continuum emission is detected to a $3\sigma$ limit of $< 2.8\ \mu$Jy at 6 GHz. Including GS~2000+25, four quiescent stellar-mass black holes with $L_X < 10^{32}$ erg s$^{-1}$ have deep simultaneous radio and X-ray observations and known distances. These sources all have radio to X-ray luminosity ratios generally consistent with, but slightly lower than, the low state radio/X-ray correlation for stellar-mass black holes with $L_X > 10^{32}$ erg s$^{-1}$. Observations of these sources tax the limits of our current X-ray and radio facilities, and new routes to black hole discovery are needed to study the lowest-luminosity black holes.

\end{abstract}


\section{Introduction} \label{sec:intro}

The relationship between radio luminosity ($L_R$) and X-ray luminosity ($L_X$) is one of the primary observational tools used for studying the connection between inflows (accretion) and outflows (jets) in accreting stellar-mass black holes \citep{Corbel13}. A strong correlation exists between $L_R$ and $L_X$ for X-ray binaries in the low/hard and quiescent states ($\lesssim 1$\% Eddington luminosity). In these states, a typical interpretation is that X-rays are produced through inverse Comptonization processes in the inner regions of a hot accretion flow, while the radio emission is synchrotron radiation originating from a relativistic jet \citep{fender04}. As a population, black holes follow a correlation L$_R \propto$ L$_X^{0.61 \pm 0.03}$ with $\sim$0.3 dex of intrinsic scatter \citep{Gallo18}. However, individual systems have been observed to follow correlations with different correlation slopes \citep{Gallo14}, some of which even change their slopes with luminosity (e.g., \citealt{Coriat11}).

Most observations of black holes in the low state were made at relatively high luminoisities, but in order to fully understand the $L_R$--$L_X$ relation, we must also study dim sources to constrain the behavior of the correlation at low luminosities ($L_X \lesssim 10^{33}$ erg s$^{-1}$). Below $L_X$ = 10$^{32}$ erg s$^{-1}$, there are only three dynamically confirmed black holes with simultaneous measurements of $L_X$ and $L_R$ and robust distance constraints \citep{Gallo06, Gallo14, Ribo17, Dincer18}.
While other black holes in this luminosity range have previously been observed with non-simultaneous observations, the strong short- and long-term variability (factors of $\gtrsim$3; e.g., \citealt{millerjones11, Dzib15, Dincer18}) in quiescence makes comparisons between the radio and X-ray unreliable.
It is also clear that individual sources show scatter about the relation when observed at different times (e.g., A0620-00, \citealt{Dincer18}), and hints of different slopes of the $L_R$--$L_X$ relation for different sources (e.g., XTE J1118+480; \citealt{Gallo14}). Needless to say, additional simultaneous deep radio and X-ray observations are required for quiescent black holes. There are only a handful of nearby ($\lesssim$3 kpc) black holes, and each one represents a precious opportunity to constrain the $L_R$--$L_X$ relation and probe accretion--jet coupling at low luminosities. 

GS~2000+25 is one such system --- a black hole X-ray binary at a distance of $2.0_{-0.4}^{+0.6}$ kpc \citep{Harlaftis96}. GS~2000+25 was discovered by the \emph{Ginga} satellite via its 1988 X-ray outburst \citep{makino88}. Using \emph{Ginga}, \citet{Tsunemi89} monitored GS~2000+25 for several weeks as the outburst began to decay, though the source clearly remained in the soft state. Pointed observations of GS~2000+25 were conducted in 1993 and 1994 with ROSAT \citep{verbunt94}, though the source was not detected down to a 1-10\,keV luminosity of $<7\times10^{31}\,$erg s$^{-1}$ (assuming a photon index of $\Gamma=2.1$ and $N_H=8\times10^{21}\,$ cm$^{-2}$; \citealt{narayan97}). 

Subsequent optical observations were used to dynamically confirm the X-ray binary as containing a black hole of $8.4\pm1.3$\,M$_{\odot}$, with a K dwarf secondary, orbiting at a period of $\sim$8.3 hr \citep{Casares95, Filippenko95, Ioannou04}.

The first quiescent detection of GS~2000+25 in X-rays was made with \emph{Chandra} Cycle 1 observations \citep{garcia01}. These data showed a significant but faint detection of the source with just five counts, yielding an X-ray luminosity of a few $\times 10^{30}$ erg s$^{-1}$. Non-simultaneous 8.46\,GHz radio continuum observations from 2009 using the pre-upgrade Very Large Array using CnB and C configurations did not detect the binary in quiescence, reaching a $3\sigma$ rms sensitivity of $30.7$ $\mu$Jy \citep{millerjones11}. These observations correspond to an upper limit on the radio spectral luminosity, $< 4.4 \times 10^{17}$ erg s$^{-1}$ Hz$^{-1}$.

This study presents the first strictly simultaneous radio and X-ray observations of GS 2000+25, using the Karl G.\ Jansky Very Large Array (VLA) and \emph{Chandra X-ray Observatory}, allowing us to accurately compare the system to the $L_R$--$L_X$ relation. In \S 2, we analyze the 2016 radio and X-ray data, and revisit the archival \emph{Chandra} data published by \citet{garcia01}. In \S 3, we discuss the implications of these observations, before briefly concluding in \S 4.

\section{Data Analysis} \label{sec:data}
\subsection{Radio Observations}
GS~2000+25 was observed for 15 hr with the VLA on two successive days: 2016 Dec 31 and 2017 Jan 1.
These two observations were each 7.5 hr long, and the 2016 Dec 31 block was entirely encompassed within our new \emph{Chandra} observations (see \S \ref{newchandra}). The VLA data were taken with C band receivers using 3-bit samplers, covering the 4--8 GHz frequency range with two 2048-MHz-wide subbands centered at 5 GHz and 7 GHz. The VLA was in its most extended A configuration, with a maximum baseline of 36.4 km. For both blocks, 3C48 was used as the bandpass and absolute flux density calibrator; it was observed at the end of each observation for 7 minutes on source. The complex gain calibrator was J2003+3034, which was observed for 1 minute between each on-source 10-minute block.

 The data were processed using the Common Astronomy Software Application \citep[CASA;][]{2007ASPC..376..127M} v5.4.0, using calibrated products provided from the VLA CASA calibration pipeline.  We concatenated the two days of data together and imaged them with CASA's {\tt clean} task. We carried out multi-frequency synthesis imaging with frequency-dependent deconvolution (nterms=2) and used Briggs weighting with a robust value of 1. The resulting image, with a central frequency of 6 GHz, has a restored synthesized beam of $0.37'' \times 0.34''$ and a central rms 
 value of 0.86 $\mu$Jy beam$^{-1}$.
 
 No significant radio continuum source is detected at the known optical location (RA= 20$^h$02$^m$49.580$^s$, Dec= 25$^{\circ}$14$^{\prime}$11.30$^{\prime\prime}$, J2000; \citealt{liu07}) of GS 2000+25. We set a $3\sigma$ upper limit of 
$< 2.6$ $\mu$Jy on the flux density of a point source at this location (see Figure 1, top). Assuming a flat spectrum, this corresponds to a $3\sigma$ radio luminosity upper limit of $< 6.2\times10^{25}\,(d/2 {\rm \,\, kpc})^2$ erg s$^{-1}$ at a reference frequency of 5 GHz.

\begin{figure}
\includegraphics[width=0.48\textwidth]{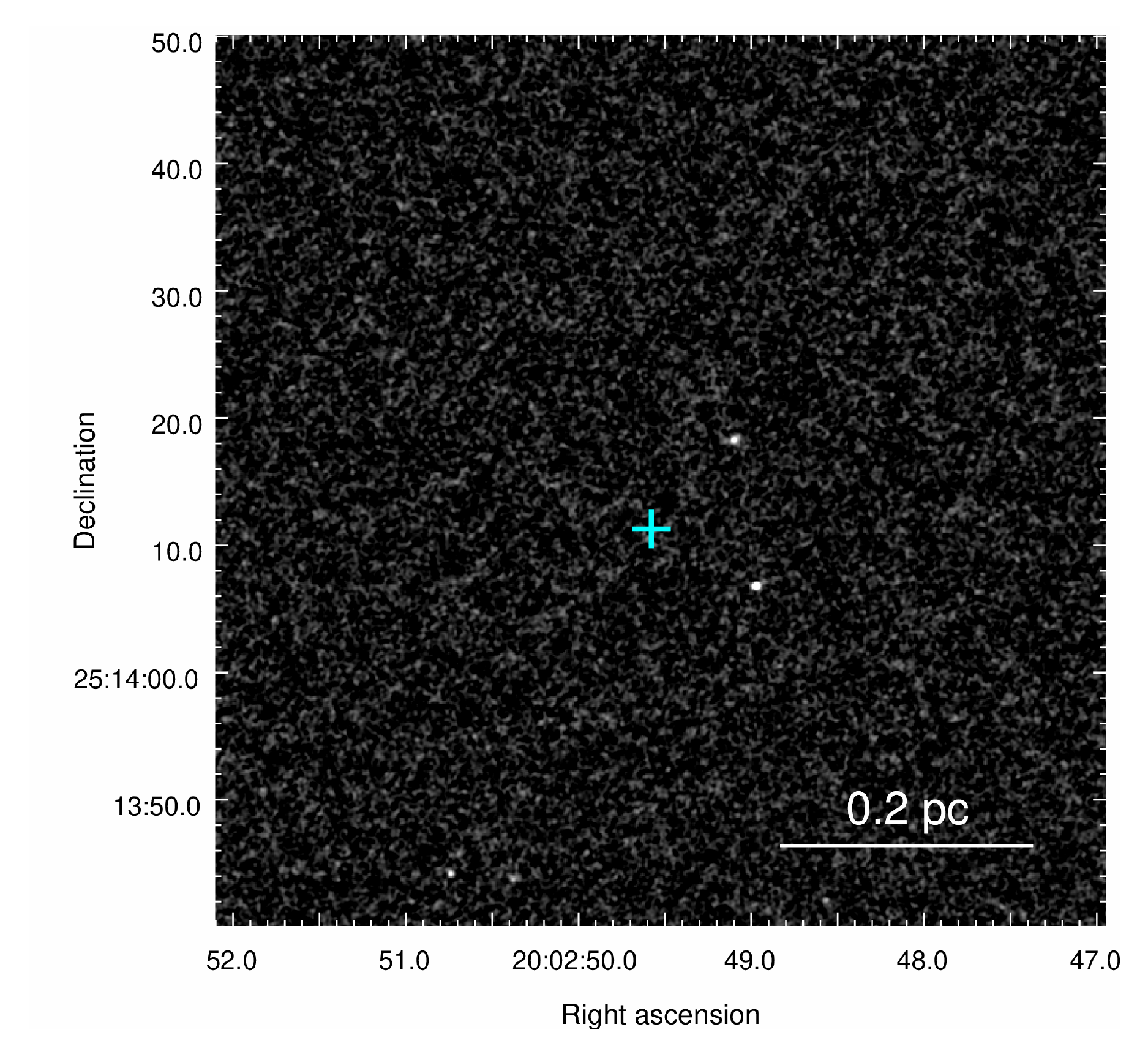}
\vspace{-0.1cm}
\includegraphics[width=0.48\textwidth]{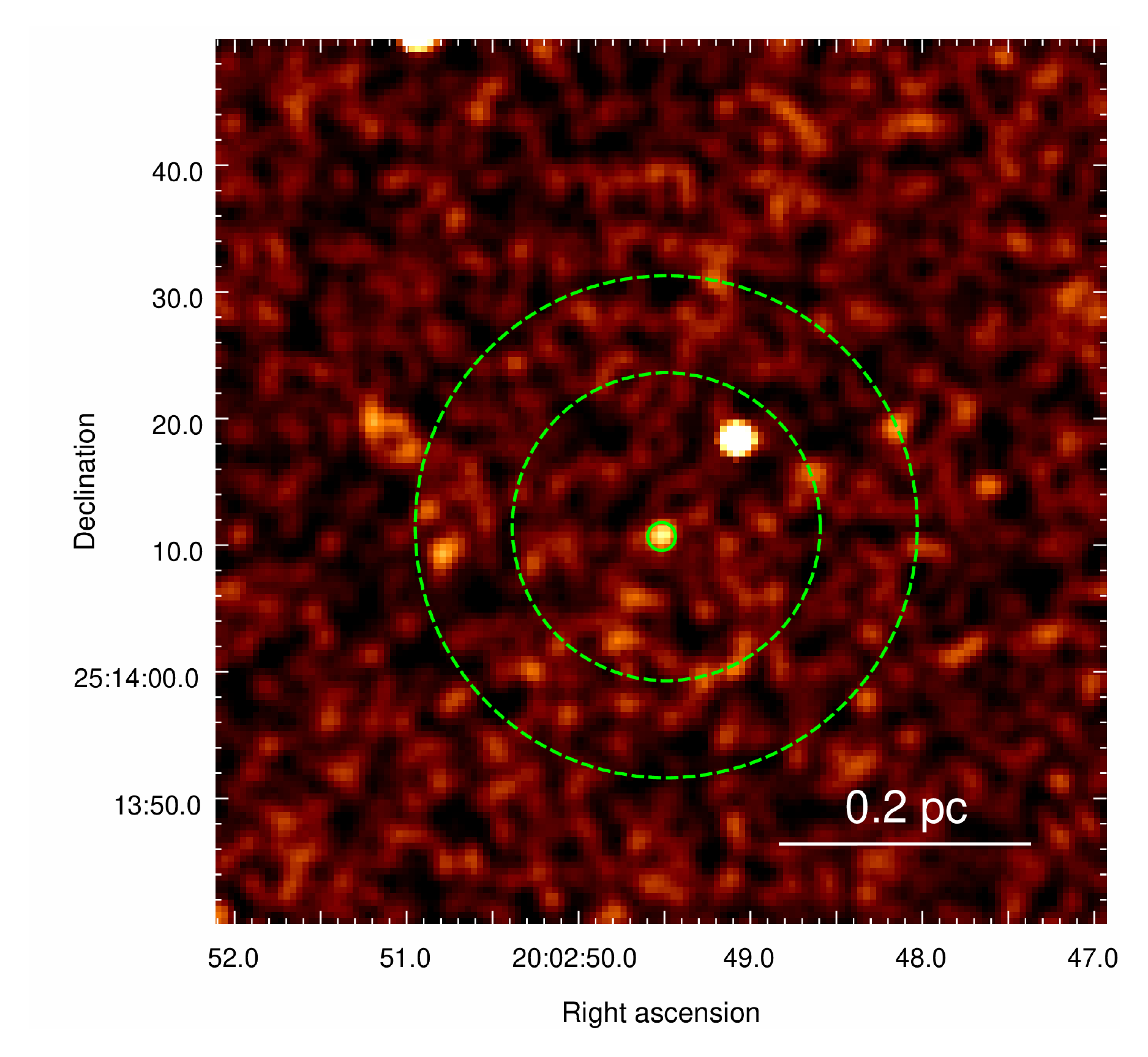}
\vspace{-0.1cm}
 \caption{Top panel: VLA 6\,GHz image of GS\,2000+25. The optical position is marked with the cyan cross. The source is not detected, with a 3$\sigma$ upper limit of $< 2.8\,\mu$Jy. Bottom panel: Stacked (1999 \& 2016) {\it Chandra} ACIS image of GS\,2000+25. For visualization purposes, the image has been adaptively smoothed using a Gaussian kernel of $\sigma=3$ pixel, with a pixel size of 0.492$^{\prime\prime}$. There is a clear detection of an X-ray source associated with the binary. The source extraction region is outlined by the solid green circle, while the dashed annulus outlines the background extraction region.}
  \label{fig:gs2000}
\end{figure}

\subsection{X-ray Observations}

\subsubsection{2016 Data}\label{newchandra}
\textit{Chandra} observed GS 2000+25 for 55.1 ks on 2016 Dec 31, starting at UT 12:05:50 (ObsID 17861) --- simultaneous with our first VLA observation for 27 ks. The target was placed at the nominal aimpoint on the S3 chip of the Advanced CCD Imaging Spectrometer \citep[ACIS;][]{garmire03}.  

Data were reduced following standard techniques in the \textit{Chandra} Interactive Analysis of Observations ({\tt CIAO}; \citealt{fruscione06}) software v4.8 and CALDB 4.7.0.   We first reprocessed the data using {\tt chandra\_repro}, and then checked that there were no time periods with anomalously high background.  Images were filtered from 0.5--7.0 keV to avoid energy ranges with higher background.

Source counts were extracted over a circular aperture (centered on the known optical position of GS 2000+25) with radius 1.1$\arcsec$, which corresponds to the 0.9 enclosed energy fraction (at 2.3 keV) at the position of the target's location on the ACIS chip.  Background counts were measured over a source-free annulus with inner and outer radii of 12$\arcsec$ and 20$\arcsec$, respectively.   Within the source aperture we find 8 counts, of which 0.4 are expected to be from the background. Using the method of \citet{kraft91}, this is a detection at the $>$99\% level (see Figure \ref{fig:gs2000}, bottom).  After dividing the net counts by 0.9 (as an aperture correction) and considering the 90\% confidence interval from \citet{kraft91}, we find a net count rate of $1.5^{+1.1}_{-0.8} \times 10^{-4}$ ct s$^{-1}$ for 0.5--7.0 keV.  

Since there are too few counts for a spectral analysis, we estimate fluxes using WebPIMMS\footnote{\url{http://cxc.harvard.edu/toolkit/pimms.jsp}} and the appropriate \textit{Chandra} Cycle 17 effective area curves.  For a column density $N_{\rm H} = 8 \times 10^{21}$ cm$^{-2}$ \citep{narayan97} and power-law spectrum with photon index $\Gamma=2.1$ \citep[typical of quiescent black hole X-ray binaries;][]{plotkin13, reynolds14}, we find absorbed and unabsorbed 1--10 keV X-ray fluxes of $1.8 \times 10^{-15}$ and $2.4 \times 10^{-15}$ erg s$^{-1}$ cm$^{-2}$, respectively. This equates to an unabsorbed luminosity of $1.1^{+1.0}_{-0.7} \times 10^{30}\,(d/2 {\rm \,\, kpc})^2$ erg s$^{-1}$.

\subsubsection{1999 Archival Data}

To ensure the most accurate comparison to our new observations, we re-analyzed the published quiescent X-ray observations from \citet{garcia01}. These were obtained on 1999 Nov 5 with \emph{Chandra}/ACIS-I and covered about 18 ks of good time (ObsID 96). Using the same analysis procedures as above, we find 5 net counts.
Using $N_{\rm H} = 8 \times 10^{21}$ cm$^{-2}$ and $\Gamma=2.1$, the unabsorbed luminosity in the 1999 observation is $2.0^{+2.0}_{-1.2} \times 10^{30} \,(d/2 {\rm \,\, kpc})^2$ erg s$^{-1}$ over the 1--10 keV band\footnote{\citet{garcia01} report luminosities over a different energy range. Converting their reported count rate to an unabsorbed 1--10 keV luminosity adopting our model parameters, our results are consistent within the uncertainties.}.

\begin{figure*}[!t]
\includegraphics[width=0.98\textwidth]{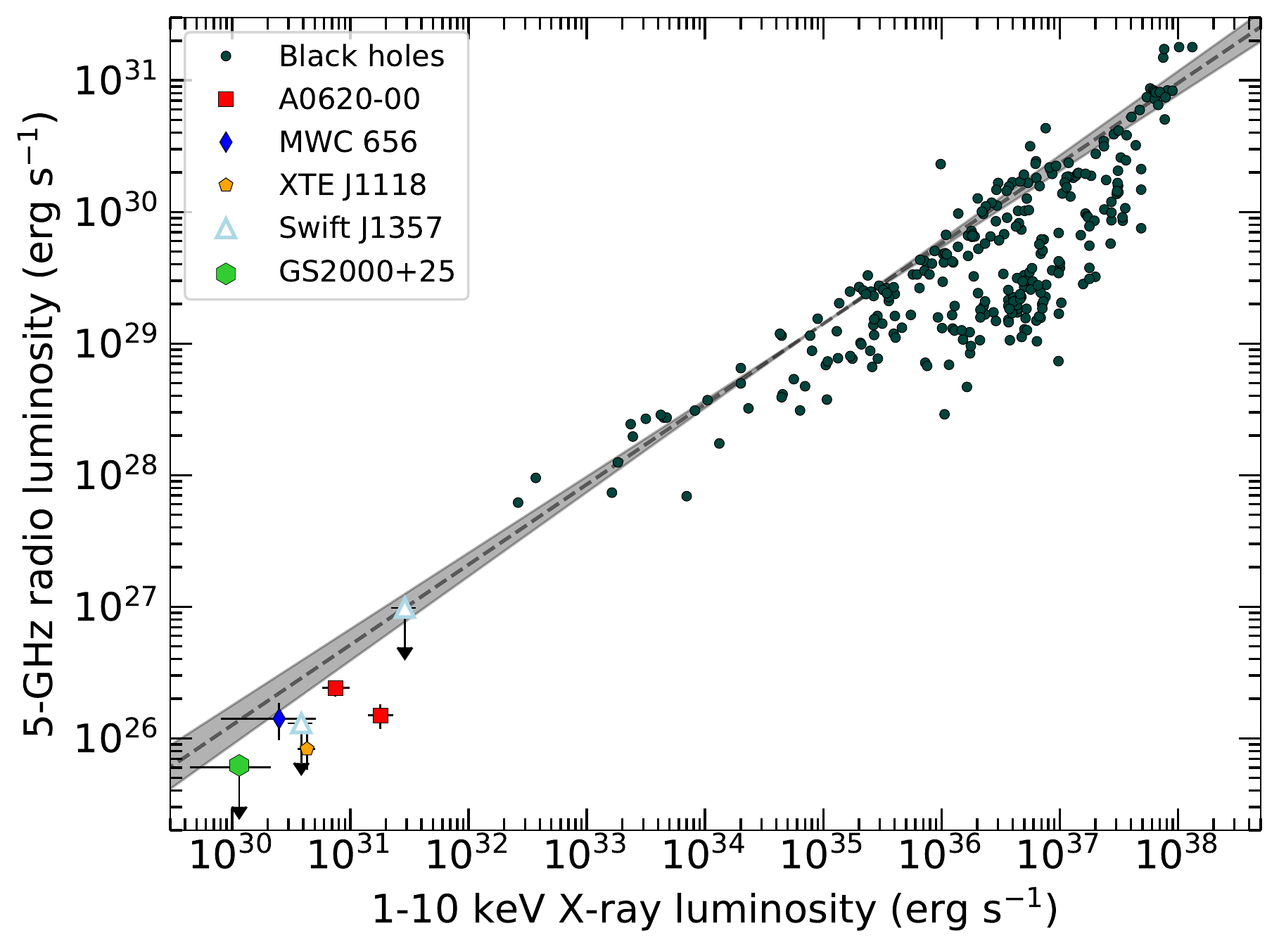}
\caption{Radio/X-ray correlation for stellar-mass black holes, with GS~2000+25 depicted as a green hexagon. Black circles indicate hard-state black holes \citep{Tetarenko16}, while sources in quiescence are individually highlighted: red squares for A0620--00 \citep{Gallo06, Dincer18},  blue diamond for MWC~656 \citep{Ribo17}, and orange pentagon for XTE~J1118+480 \citep{Gallo14}. Due to the large uncertainty on the distance (2.3 -- 6.3\,kpc), we plot the luminosity range derived from the distance extremes for Swift J1357.2-0933 (empty cyan triangles; \citealt{Plotkin16}), though the distance is likely to be $\gtrsim6$\,kpc \citep{Charles19}. The dashed line represents the best-fit radio/X-ray relation for black holes in the low/hard state from \citet{Gallo14}, with the 1$\sigma$ error expressed by the shaded region.}
  \label{fig:lrlx}
\end{figure*}

\subsection{Distance constraints}

A challenge in precisely placing low-mass X-ray binaries
on the radio/X-ray correlation is that their distances are often uncertain. Here we briefly evaluate distance estimates for GS\,2000+25.

So far in this work, we have used the distance of \citet{Harlaftis96}, $d=2.0_{-0.4}^{+0.6}$ kpc. This measurement comes from a combination of measurements of the GS\,2000+25 secondary: its $K$ band apparent magnitude, effective temperature, and projected size (as determined by Roche geometry). The uncertainty is principally determined by the poorly-constrained inclination of the binary. Relative to distance measurements determined using the optical ($V$) magnitude, infrared measurements are less sensitive to line-of-sight extinction and contamination from the accretion disk.

We can obtain an independent estimate of the distance by comparing the extinction of GS\,2000+25 to three-dimensional reddening maps, which are of much higher quality than those available when the main papers were being written on this source two decades ago. We use the correlation between galactic extinction and the equivalent width of the Na interstellar absorption lines, $\lambda$5890\,\AA\ and $\lambda$5896\,\AA\ \citep{Poznanski12}. Using the unresolved Na doublet equivalent width for GS\,2000+25 reported in \citet{Charles91}, we find an interstellar extinction of $A_V=1.11\pm0.25$ mag. Comparing this extinction to two sets of three-dimensional maps, which combine \emph{Gaia}, Pan-STARRS, and 2MASS \citep{Green19} or \emph{Gaia}, 2MASS, and WISE  \citep{Chen19}, we find distances of 0.73$\pm$0.05 kpc and $1.0_{-0.2}^{+0.6}$ kpc, respectively. The small uncertainty in the \citet{Green19} estimate arises from a dust cloud in the model at $\sim 0.75$ kpc, beyond which the modeled extinction rises to $A_V \gtrsim 2$ mag.

We can also use the relation between hydrogen column density ($n_{\mathrm H}$) and extinction \citep{Bahramian15} to estimate the distance. A value of $n_{\mathrm H}=8\times10^{21}$ cm$^{-2}$ \citep{narayan97} implies $A_V=2.85$ mag and, hence, $d=2.0_{-0.6}^{+0.9}$ kpc (based on the \citealt{Chen19} reddening maps; the value for the \citet{Green19} maps is similar).

The discrepancy between these optical and X-ray extinction estimates (and corresponding distance measurements) is substantial, and indeed \citet{Charles91} mentioned that different interstellar lines in their optical spectra of GS\,2000+25 implied inconsistent values for the extinction. The original \citet{Harlaftis96} distance assumed $E(B-V) = 1.5$ and would be slightly (around 10\%) larger if a lower reddening of $E(B-V) = 0.9$ were adopted, but this source of uncertainty is small compared to that in the system inclination, and in any event would not change our conclusions.

While not conclusive, there is agreement on a distance of $\sim 2$ kpc from modeling the infrared photometry of the secondary and the extinction implied by the X-ray data. The optical extinction may indicate a somewhat nearer distance. In any case, a distance meaningfully in excess of 2--3 kpc seems unlikely based on the available data, and thus we are confident the X-ray and radio luminosities (or limits) that we report below for GS\,2000+25 are not substantial underestimates. The source might well be bright enough that a geometric parallax could be determined in a future \emph{Gaia} data release.

\section{Results and Discussion}

GS~2000+25 is the lowest-luminosity black hole with high-quality, simultaneous observations at both radio and X-ray wavelengths. Our upper limit on its radio luminosity is the deepest to date, constraining the spectral luminosity to be $<$1.2$\times10^{16}$ erg s$^{-1}$ Hz$^{-1}$ (for a distance of 2.0 kpc; \citealt{Harlaftis96}). Our 2016 X-ray observations measure a 1--10 keV luminosity of $1.1^{+1.0}_{-0.7} \times 10^{30}$ erg s$^{-1}$, which corresponds to an X-ray Eddington ratio of $\sim 10^{-9}$.
Comparing our new {\it Chandra} observation to the re-analyzed Cycle 1 archival data, we find marginal evidence of X-ray variability. Between 1999 and 2016, the 1--10 keV unabsorbed flux of GS~2000+25 decreased by a factor of two. However, due to the low number of counts in each observation, and hence large uncertainties, we cannot conclusively confirm the presence of any variability.

Our 2016 measurements are the only simultaneous radio and X-ray observations of GS~2000+25 in the low/hard or quiescent states. For the two months following its discovery in X-ray outburst (May--June 1988; \citealt{Tsunemi89}), GS~2000+25 was monitored with radio observations (1.5--15 GHz) by \citet{Hjellming88}. During this time, it appeared as a fading synchrotron source with a radio spectrum, $S_{\nu} \propto \nu^{-0.5}$. However,  the X-ray spectrum was in the soft state throughout this period \citep{Efremov89}, and so these X-ray and radio observations are not appropriate for placement on the $L_{R}-L_{X}$ plane.

We place GS~2000+25 on the $L_R$--$L_X$ plane in Figure \ref{fig:lrlx}, and  compare it with other stellar-mass black holes in the low/hard and quiescent states. Radio luminosities for all black holes are normalized to 5 GHz assuming a flat spectrum ($S_{\nu} \propto \nu^0$). The black dashed line represents:
\begin{equation}
L_R = 4.4 \times 10^7\ L_X^{0.61}\,\,\mathrm{erg\,s}^{-1}
\end{equation}
as fit by \citet{Gallo14} to 24 black hole X-ray binaries (essentially all of which have $L_X > 10^{33}$ erg s$^{-1}$).

GS~2000+25 is the fourth member of a growing population of low-luminosity quiescent black holes that constrain the $L_R$--$L_X$ relation at low luminosities ($L_X < 10^{32}$ erg s$^{-1}$), joining XTE~J1118+480 \citep{Gallo14}, MWC~656 \citep{Ribo17}, and A0620--00 \citep{Gallo06,Dincer18} at this low-luminosity extreme. A fifth source, Swift J1357.2-0933, also has strictly simultaneous and deep radio and X-ray observations in quiescence \citep{Plotkin16}, but we do not include it because of its very uncertain distance.

It is noteworthy that all four of these quiescent black holes are below the \citet{Gallo14} $L_R$--$L_X$ relation (if only slightly; see Figure \ref{fig:lrlx}), which is mostly determined by the radio-luminous branch of black hole X-ray binaries with $L_X > 10^{32}$ erg s$^{-1}$. Since the number of sources in this region is still small, one interpretation of Figure \ref{fig:lrlx} is that, due to intrinsic dispersion or distance uncertainties, these quiescent sources happen to sit below the mean relation by chance (since $L_R$--$L_X$ is nonlinear, distance uncertainty vectors are not parallel to the relation; see e.g., \citealt{millerjones11}). Beyond the realm of random uncertainties, it is plausible that black hole binaries could vary in the normalization or slope of their individual $L_R$--$L_X$ relations. As an example, \citet{Gallo14} argue that XTE~J1118+480 shows a marginally steeper $L_R$--$L_X$ relation compared to the full sample of black holes.

Another possibility is that there is a true change in the nature of the $L_R$--$L_X$ relation at $L_X \lesssim 10^{-7} L_{\rm edd}$. If such a change is present, it is much less extreme than implied by some scenarios, such as a transition in X-ray emission from accretion flow-dominated to jet-dominated (and an associated strong steepening of the $L_R$--$L_X$ relation; e.g., \citealt{yuan05}, who suggest $L_R \sim L_X^{1.23}$). Instead, a slightly steeper exponent (compared to $L_R \sim L_X^{0.61}$ in Figure \ref{fig:lrlx}) might better fit the quiescent stellar-mass black holes in the low-Eddington regime.
The evidence for a slope change is weak with the current observational constraints, and substantially more data would be necessary to investigate this possibility.

\section{Conclusions and Future Work}
Using deep, simultaneous radio and X-ray observations, we have placed GS~2000+25 on the $L_R$--$L_X$ plane. GS~2000+25 was not detected using the VLA, placing a 3$\sigma$ upper limit of $< 6.2 \times 10^{25}\,(d/2 {\rm \,\, kpc})^2$ erg s$^{-1}$ on the C-band radio luminosity. We detected GS~2000+25 with \emph{Chandra} at an unabsorbed X-ray luminosity of $1.1^{+1.0}_{-0.7} \times 10^{30}\,(d/2 {\rm \,\, kpc})^2$ erg s$^{-1}$. Therefore, GS~2000+25 is the least luminous black hole X-ray binary known \citep{Gallo08}.

A number of recent papers have made significant efforts to fit the broadband spectral energy distributions of quiescent black holes, in an attempt to distinguish the different sources of emission in the system, including the donor, accretion flow, and jet (e.g., \citealt{plotkin15, Dincer18}). Such studies have great difficulty in distinguishing among emission models owing to, among other factors, the faintness of the sources in X-ray and radio bands.
Short of waiting for the next generation of observational facilities, it is critical to find new black hole X-ray binaries systems to study. Since new black hole outbursts are discovered rarely, and even fewer of these systems are nearby, there is a crucial need for surveys that can discover such systems in quiescence. Black holes can be detected and differentiated from other source classes using their $L_R$ and $L_X$ \citep[e.g.,][]{Chomiuk13, Tetarenko16_m15}. Combining new radio surveys (such as the VLA Sky Survey and EMU survey with ASKAP; \citealt{norris11, lacy19}) with new X-ray surveys (e.g., eROSITA; \citealt{merloni12}) is a promising way forward over the next decade. Such joint radio/X-ray campaigns will compliment ongoing efforts to detect quiescent black hole binaries, like narrow-band emission line imaging \citep{Casares18} and X-ray surveys \citep{Jonker14}.\\

We thank an anonymous referee for helpful comments that substantially improved the paper. Support for this work was provided by the National Aeronautics and Space Administration through Chandra Award Numbers GO6-17045X, G06-17040X, GO7-18032A, and GO8-19122X issued by the Chandra X-ray Center, which is operated by the Smithsonian Astrophysical Observatory for and on behalf of the National Aeronautics Space Administration under contract NAS8-03060. JS acknowledges support from a Packard fellowship. JCAM-J is the recipient of an Australian Research Council Future Fellowship (FT140101082), funded by the Australian government. GRS acknowledges support from an NSERC Discovery Grant (RGPIN-06569-2016).

The scientific results reported in this article are based on observations made by the Chandra X-ray Observatory. The National Radio Astronomy Observatory is a facility of the National Science Foundation operated under cooperative agreement by Associated Universities, Inc. This research has made use of software provided by the Chandra X-ray Center (CXC) in the application package CIAO.

\facilities{CXO, VLA}

\bibliography{ref}



\end{document}